\documentstyle[aps,prl,multicol]{revtex}
\input{psfig}
\begin{document}
\renewcommand{\[}{\begin{equation}} 
\renewcommand{\]}{\end{equation}}
\newcommand{\ie}{{\it i.e.}}
\title{Effect of screening on shot noise in diffusive mesoscopic 
conductors}
\author{Y. Naveh\cite{email}, D. V. Averin, and K. K. Likharev}
\address{Department of Physics, State University 
of New York, Stony Brook, NY 11794-3800}
\date{\today }
\maketitle

\begin{abstract}
  Shot noise in diffusive mesoscopic conductors, at finite observation
  frequencies $\omega $ (comparable to the reciprocal Thouless time
  $\tau _T^{-1}$), is analyzed with an account of screening.  At low
  frequencies, the well-known result $S_I(\omega)=2eI/3$ is recovered.
  This result is valid at arbitrary $\omega \tau_T$ for wide
  conductors longer than the screening length. However, at least for
  two very different systems, namely, wide and short conductors, and
  thin conductors over a close ground plane, noise approaches a
  different fundamental level, $S_I(\omega) = eI$, at $\omega \tau
  _T\gg 1$.
\end{abstract}

\begin{multicols}{2}

The study of non-equilibrium fluctuations of current (''shot noise'')
provides important information about microscopic transport properties
of conductors. This explains the recent interest in shot noise in
various mesoscopic systems (see, {\it e.g.}, Ref. \cite{de Jong 96}).
In the case of a diffusive mesoscopic conductor ($l\ll L\ll l_{{\rm
in}}$, where $L$ is sample's length, while $l$ and $l_{{\rm in}}$ are
the elastic and inelastic scattering lengths, respectively), the
low-frequency limit of the shot noise spectral density $S_I(\omega )$
was found to be 1/3 of the classical Schottky value $2eI$, where $I$
is the average current through the sample \cite{Beenakker 92,Nagaev
92,Nazarov 94}.

Most theoretical works on this subject did not discuss
screening at all. However, screening, at least in the external
electrodes, has to be present in the problem because
of the very definition of the noise current (see, {\it e.g.}, the
remarks by Landauer \cite{Landauer 96}): although the fluctuations
leading to the shot noise originate in the conductor itself, the
observable noise is that of the current $I^e(t)$ induced by these
fluctuations in the external circuit.  With a permissible
simplification, $I^e(t)$ can be considered as a current in
semi-infinite electrodes, but {\it deep} inside them, at distances
much larger not only than $l$ and $l_{{\rm in}}$, but also than the
screening length from the interface with the conductor.

The purpose of this work was to analyze the frequency dependence of
the shot noise in diffusive conductors due to the finite ``Thouless''
time $\tau_T$ of electron diffusion along the sample\cite{Buttiker
96}. We assumed here that the applied voltage is large enough, $eV \gg
\hbar/\tau_T, k_B T$ so that in the relevant frequency range $\omega
\sim \tau_T^{-1}$ we can neglect quantum noise appearing at $\omega
\sim eV/\hbar$ \cite{Altshuler 94}. We considered a dirty (metallic or
semiconducting) conductor which connects two identical electrodes with
screening length $\lambda _e$ generally different from that of the
conductor ($\lambda )$, within two analytically solvable models. In
the first model [Fig. 1(a)] the conductor is assumed to be short and
thick ($L\ll t$, where $t$ is the thickness, \ie, the smallest
transversal size).  In the second case [Fig.  1(b)] we consider a thin
conductor located close to a well-conducting ground plane: $t,d\ll L$
(in this case the conductor width $W$ is arbitrary).  In both cases we
assume that the Fermi level of the conductor is close to that of the
electrodes even before they have been brought into contact, so that
there are no Schottky barriers at their interfaces (see inset in
Fig.~1a). In semiconductor samples, this balance can be readily
achieved by either the appropriate doping or electrostatic gating. The
interfaces between the electrodes and the conductor are assumed to be
smooth (adiabatic) on the scale of the elastic mean free path $l$, but
sharper than $\lambda $ and $L$. Thus, the complete hierarchy of
length scales assumed in this work is $ \lambda _F\ll l\ll \Delta x\ll
\lambda ,\lambda _e,L\ll l_{{\rm in}}, $ where $\lambda _F$ is the
Fermi surface wavelength in the conductor and $\Delta x$ characterizes
the interface smoothness.

Following Nagaev\cite{Nagaev 92}, we start with the
semiclassical Boltzmann-Langevin equation\cite{Kogan 69} (justified by
the assumption $ \lambda _F\ll l$). Integration of this equation over
the electron momenta leads to a ''drift--diffusion-Langevin'' equation
for the current density fluctuations
\[
\label{generalj}{\bf j}({\bf r},t)=\sigma ({\bf r}){\bf E}({\bf 
r},t)-D({\bf %
r}){\bf \nabla }\rho ({\bf r},t)+{\bf j}^s({\bf r},t), 
\]
which is valid even if parameters of the conductor are changing in
space (on the scale $\Delta x\gg l$). Here $\rho ({\bf r},t)$ and
${\bf E}({\bf r},t)$ are the local fluctuations of charge density and
electric field, respectively, and ${\bf j}^s({\bf r},t)$ represents
the Langevin fluctuation sources of current:
\[
\label{intsources}{\bf j}^s({\bf r},t)=e\tau ({\bf r})\sum_{{\bf 
k}}{\bf v}_{%
{\bf k}}{J}^s({\bf r},{\bf k},t), 
\]
\noindent 
\begin{figure}[tb]
\centerline{\hspace{-31pt} \psfig{figure=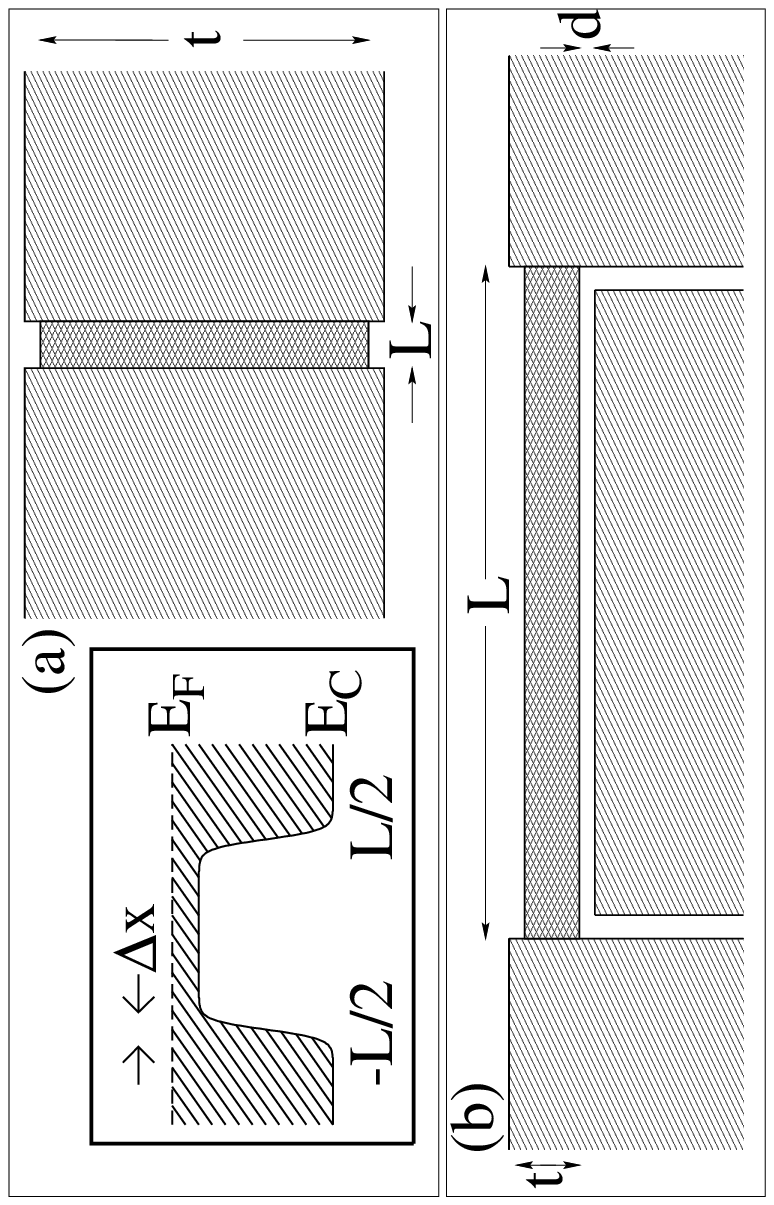,angle=-90,width=80mm}}
\narrowtext
\vspace{0.5cm}
\caption{Two geometries studied in this work (schematically): (a) -
short and wide conductor; (b) - thin and long conductor close to a
ground plane. The inset shows the conduction band edge diagram of the
structures.}
\label{1model}
\end{figure}
\noindent
with ${\bf v}_{{\bf k}}$ the velocity of an electron and $\hbar {\bf
  k}$ its momentum. The correlator for the random sources ${J}^s({\bf
  r},{\bf k},t)$ has been found in \cite{Kogan 69} (see also
\cite{Nagaev 92}). In equations (\ref{generalj}) and
(\ref{intsources}), $\tau ({\bf r})$,{\bf \ }$\sigma ( {\bf r})$, and
$D({\bf r})$ are the local elastic scattering time, conductivity, and
diffusion constant, respectively.  Equation~(\ref{generalj}) should be
solved together with the usual Poisson and continuity equations.

For the first model [Fig. 1(a)], we can use 1D versions of all these
equations along the dc current direction ($x$), with all the
variables integrated over the cross-section $A$ of the conductor.
In this case, the combination of the Poisson and continuity equations,
integrated over $x$, provides a simple relation between the Fourier
images of current and longitudinal electric field:
\[
\label{integrated}I_\omega (x)=\frac{i\omega \varepsilon 
(x)}{4\pi }E_\omega
(x)+I_{\omega ,}^e 
\]
where $\varepsilon (x)$ is the local dielectric constant, while
$I_\omega ^e$ is an integration constant which has the physical sense
of current fluctuations induced deep inside the electrodes (where
$E_\omega =0$). This constant can be found from the condition that the
current fluctuations do not affect the voltage applied to the
structure:
\[
\label{fixedvoltage}\delta V_\omega =-\int_{-\infty }^\infty 
dx\,E_\omega
(x)=0. 
\]

In our second model [Fig. 1(b)] the electrostatic potential $\Phi
_\omega (x)$ is
completely determined by the local linear density $q_\omega (x)$ of
electric charge in the conductor:
\[
\label{capacitance}\Phi _\omega (x)=q_\omega (x)/C_0, 
\]
where $C_0$ is the capacitance per unit length. The justification of
this equation will be given below. 

Combining the 1D equation for current fluctuations with
Eq.~(\ref{integrated}) (first model), or with Eq.~(\ref{capacitance})
and the continuity equation (second model), we get the same simple
differential equation
\[
\label{diffeq}\frac{d^2I_\omega (x)}{dx^2}-\kappa ^2(x,\omega 
)I_\omega (x)= \frac{i\omega }{D^{\prime }(x )}I_\omega ^s(x)-\frac
1{\Lambda ^2(x)} I_\omega ^e,
\]
where $\kappa (x,\omega )=[\Lambda ^{-2}(x)-i\omega
/D^{^{\prime}}(x)]^{1/2} $ is the effective screening parameter. The
effective static screening length $\Lambda $ and diffusion constant
$D^{^{\prime}}$ are, however, different for our two models. In the
first model [Fig. 1(a)], $D^{^{\prime }}(x)=D(x)$ and $\Lambda
(x)=\lambda (x)$, where $\lambda(x) = \sqrt{\varepsilon(x) D(x) / 4\pi
\sigma(x)}$ is the usual static screening length.

In the second model, however, the effective parameters are 
renormalized: 
\[
D^{^{\prime }}(x)=D(x)+ \frac{\sigma (x) A}{C_0}, \, \, \, \, \Lambda
(x)=\infty .
\]
The reason for the renormalization of the diffusion constant $D$ is
that now the full electro-chemical potential (not only the chemical
potential) is proportional to the density $\rho_\omega$.  The second
equation is due to the fact that in this model the electric field
created by charge fluctuations is transversal (directed toward the 
ground plane) and does not lead to accumulation of electrostatic
potential along the sample.
Equation~(\ref{diffeq}) shows that spatial harmonics with $k \gg
|\kappa| = \sqrt{\omega/D'}$ contribute negligibly to the current
fluctuations $I_\omega$. Thus, at frequencies $\omega \sim \tau_T^{-1}
\ll D'/d^2, D'/t^2$, we can consider only the wavevectors $k$ (in $x$
direction) which are much smaller than $d^{-1}$, $t^{-1}$. For these
harmonics, transversal gradients of all variables dominate, leading
immediately to Eq.~(\ref{capacitance}).

In the case of the first model, equation~(\ref{diffeq}) is applicable
to both the conductor and the electrodes. The boundary conditions at
the interfaces ($x=\pm L/2$) can be derived from the continuity of the
current and the electron distribution function at the interfaces.
Integrating the latter condition over the electron momenta, and using
the continuity equation, we obtain that $I_\omega $ and $\lambda
^2dI_\omega /dx$ should be continuous across the interfaces.

Let us consider the most natural case of well-conducting
electrodes of macroscopic size ($\gg \lambda _e,l_{_{{\rm in}}}$),
with a resistance negligible in comparison with that of the conductor.
Scattering in the bulk of such electrodes does not produce
considerable current noise (see, {\it e.g.,} Sec.~4.2 of \cite{de Jong
  96}). The inelastic scattering of nonequilibrium electrons arriving
from the conductor also gives negligible noise sources, since the
number of scattering events per transferred electron in a dirty
conductor is a factor of $(L/l)^2\gg 1$ larger than the number of
inelastic collisions leading to its thermalization in the electrodes.
Therefore, we can neglect the noise sources inside the electrodes, and
the solution to Eq.~(\ref{diffeq}) can be presented in the form
\[
\label{jresponse}I_\omega (x)=\frac 1L\int_{-\frac L2}^{\frac 
L2}K_\omega
(x,x^{\prime })I_\omega ^s(x^{\prime })\,dx^{\prime }. 
\]

In the case when the conductor and electrodes are uniform along 
their length
(generally, with different $\lambda $), the kernel $%
K_\omega (x,x^{\prime })$ can be expressed with an analytical, though
bulky, formula. However, its value at $x=\pm \infty $, giving the
current in the external circuit, has a compact form
\[
\label{kinfty}K_\omega ^e(x^{\prime }) \equiv K_\omega(\pm \infty, x')
= \frac{1+{\cal A}(\omega )\cosh (\kappa x^{\prime })}{1+{\cal A}(\omega
  )[\sinh (u)/u]}
\]
where ${\cal A}(\omega )\equiv (k^2-1)/[\cosh (u)+(\lambda 
/\lambda
_e)k\sinh (u)]$, $k\equiv \kappa \lambda$, $u \equiv \kappa L/2$, 
and $\kappa =\kappa
(\omega )$ is the effective screening parameter inside the 
conductor.

Figures~2(a,b) show the distribution of the charge created by a
single, localized, low-frequency current fluctuation source, for the
case when the screening in the electrodes is strong, $\lambda _e\ll
L,\lambda$.  One can see that the total charge created by the
fluctuation source inside the conductor is compensated by the charge
layers of opposite polarity at the conductor-to-electrode interfaces.
Equation~(\ref{kinfty}) shows that at low frequencies the external
response function is uniform, $K_0^e(x^{\prime }) = 1$.
\noindent
\begin{figure}[tb]
\vspace{-80pt}
\centerline{\hspace{3pt} \psfig{figure=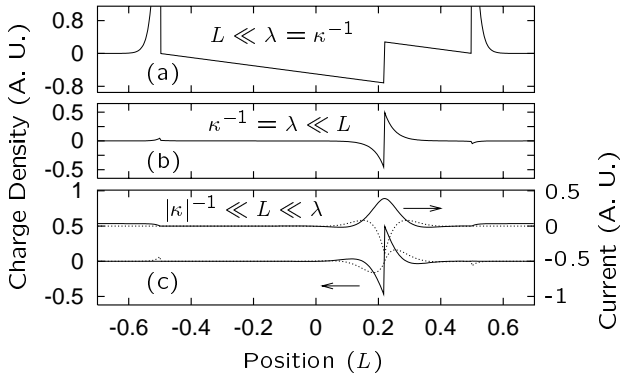,angle=-90,width=120mm}}
\narrowtext
\vspace{-32pt}
\caption{Charge density induced by a point current source at $x=0.22$
for the cases of strong screening in the electrodes ($L/\lambda _e =
70$): (a) - low frequency, no internal screening ($L/\Lambda
\rightarrow 0$), (b) - low frequency, strong internal screening
$(L/\Lambda =30)$, and (c) - high frequency $(|\kappa |L=30)$, no
internal static screening.  Solid (dotted) lines show the real
(imaginary) parts of $\rho_\omega$ and $I_\omega$. These results are
valid for both our models, but in the case of conductor over a ground
plane [Fig.~1(b)] there are no charge layers at the interfaces, and
$\Lambda$ is always infinite.}
\label{2charge}
\end{figure}
\noindent

In the limit of strong internal screening ($\lambda \ll L$), the local
electric dipole formed by the point current fluctuator is screened to
a size of the order of $ \sim \lambda $, much smaller than the
conductor length [Fig.  2(b)]. However, the external current does not
vanish and is, moreover, again independent of the fluctuator position.
This result stems from the fact that fluctuations of {\it current}
(rather than {\it charge}) do not violate the sample
electro-neutrality per se \cite{Pines 66}, and are not the subject of
electrostatic screening at distances of the order of $\lambda .$

Figure 2(c) shows the distribution of the induced charge and current
at high frequencies when $|{\rm \kappa }|L\approx (\omega \tau
_T)^{1/2}\gg 1$. Again in the most of the conductor the charge is
zero, and the only substantial difference from the case of strong
static screening is that the induced current is not constant (which
accounts for the finite displacement current).

In the case of our second model shown in Fig. 1(b) the boundary
conditions for Eq.~(\ref{diffeq}) generally depend on the exact shape
of electrodes. However, in the case of "perfect" electrodes
($\lambda_e \ll \lambda, L$) the boundary conditions are reduced
to a simple form
\[
\label{simpleBC}\rho_\omega = 0 \, \, \, \, {\rm at} \, \,\, \, x=\pm L/2
\]
(in this limit these boundary conditions are also valid for our first
model).  Equation~(\ref{simpleBC}) together with
Eq.~(\ref{capacitance}) make the condition ~(\ref{fixedvoltage})
satisfied automatically, and all the analysis is reduced to the volume
of the conductor.

Solution of Eq.~(\ref{diffeq}) with the boundary
conditions~(\ref{simpleBC}) show that the distribution of charge (but
not electric potential!) along the conductor in the second model is
generally similar to that in the first model (Fig. 2) in the
corresponding limit ($\lambda_e \ll L \ll \Lambda$). The only
difference is that the conductor over the ground plane does not induce
the thin charged layers at the conductor-to-electrode interfaces. The
physical reason of this difference is that the charge accumulated in
the conductor is now compensated by the equal and opposite charge of
the ground plane.  Oscillations of this charge are responsible for the
fact that at finite frequencies the currents through the interfaces
contain not only the "transport" (symmetric) component $I^e$, but also
an asymmetric component providing for the re-charging of the sample.
As a result the response functions for the left and right interfaces
are different:
 \[ \label{kgrplane}
K_{\omega}^\pm(x^{\prime }) = 2u
\frac{\cosh(u \pm \kappa x')}{\sinh(2u)}.
\]

In order to find the spectral density of the current fluctuations
$I^s(x,t)$ we need to know the explicit correlation function of the
fluctuation sources. In the non-equilibrium limit ($eV/2 \gg k_BT$) this
function can be readily obtained by combining Eq.~(\ref{intsources})
above with Eqs.~(10) and (14) of Ref.~\cite{Nagaev 92}:
\[
\label{corrfunc}\langle I^s(x,t)I^s(x^{\prime },0)\rangle 
=\frac{eIL}2\left(
1-\frac{4x^2}{L^2}\right) \delta (x-x^{\prime })\delta (t), 
\]
$I$ being the time-averaged current.
From Eqs.~(\ref{jresponse}) and (\ref{corrfunc}), the spectral density
of fluctuations of the current $I(x,t)$ flowing in an arbitrary
cross-section of the conductor is
\begin{eqnarray}
\label{generalnoise} \nonumber 
S_I(x,\omega ) & \equiv & 2\int_{-\infty }^\infty 
\langle
I(x,t)I(x,0)\rangle \exp (i\omega t)dt \\ 
& = & \frac{eI}L\int_{-\frac 
L2}^{\frac L2%
}|K_\omega (x,x^{\prime })|^2\left( 1-\frac{4x^{\prime 
}{}^2}{L^2}\right)
\,dx^{\prime }. 
\end{eqnarray}

For our first model, the spectral density $S_I(\omega )$ of the noise
current in electrodes can be found as $S_I(\pm \infty ,\omega ).$
For low frequencies we have $K_0^e(x^{\prime }) = 1$, confirming the
earlier result $S_I(0)=(1/3)\times 2eI$ regardless of the screening
properties of the system. At finite frequencies the shot noise depends
on the effective screening lengths of the conductor and electrodes. In
the high-frequency limit its spectral density is given by
\[
\label{highfreq}S_I(\omega )=\frac{eI}3\frac{2(\alpha 
L)^2+3}{(\alpha
L)^2+\beta L+1} \, \, \, \, \, \, {\rm at} \, \, \, \, \, \, 
\omega
\tau _T\gg 1,
\]
with $\alpha ^2\equiv 1/(2\lambda_e)^2+ 1/(\lambda_0 \lambda)
+2/\lambda_0^2$, $ \beta \equiv 1/\lambda _e+2/\lambda_0$, and
$\lambda_0\equiv (2\lambda )^2 \sqrt{\omega/2D}$. This dependence is
shown in Fig.~3(a).  In the case of strong screening (either $\lambda
\ll L,$ or $\lambda _e\ll L, $ or both), the noise suppression factor
equals 1/3 even at high frequencies. For intermediate screening
($\lambda \sim L$ or $\lambda_e\sim L$) the suppression factor can be
as small as 1/5. However, if both effective screening lengths are much
larger than $L$, the high frequency noise assumes the value of
$(1/2)\times 2eI$, with a crossover near 
$1/\tau _T$ -- see Fig.~3(b). 
\noindent
\begin{figure}[tb]
\vspace{-70pt}
\centerline{\hspace{-150pt} \psfig{figure=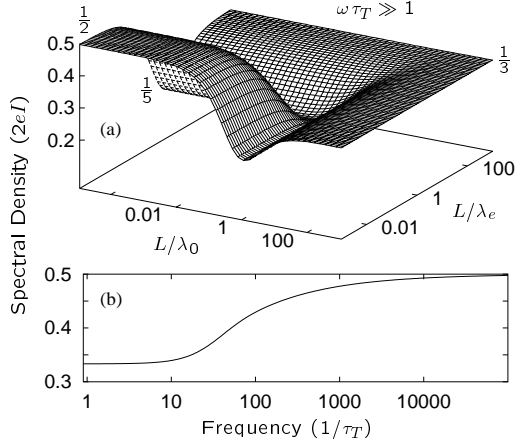,angle=0,width=150mm}}
\narrowtext
\vspace{-225pt}
\caption{(a) - High frequency shot noise in the external circuit as a
function of $ L/\lambda_0 $ and $L / \lambda _e$.
(b) - Frequency dependence of the shot noise, which is valid for short wide
conductors at $L \ll \lambda, \lambda_e$, and for thin conductors over a
ground plane for $L \gg \lambda_e$, but arbitrary $\lambda$.}
\label{3noise}
\end{figure}
\noindent

The last result can be interpreted as follows. At high frequencies
only the fluctuators near the interface induce noise current in the
electrodes (the response functions $K_\omega^e(x')$ exponentially
decrease far from the interface), so that each edge of the conductor
can be regarded as an independent noise source. Since the electron gas
on one side of this "point contact" is strongly overheated, its noise
should obey the classical Schottky formula. The elementary addition of
these two independent noise sources with equal effective resistances (see,
{\it e.g.},  Ref. \cite{Beenakker 92}) gives the factor (1/2).

In the ground-plane model, assuming a natural measurement scheme with
an instrument symmetric with respect to the ground plane, we accept
$I_\omega^e = [I_\omega(-L/2) + I_\omega(L/2)]/2$, thus excluding the
asymmetric component of the noise current, responsible for re-charging
of the conductor. In this case, $S_I(\omega)$ is given by the same
result as for the first model in the limit $\lambda, \lambda_e \gg L$
[Fig.~3(b)] [this can be verified directly by comparing
Eq.~(\ref{kgrplane}) with Eq.~(\ref{kinfty})\cite{Nagaev 97}]. Notice,
however, that now this limitation is {\it not} valid, and a
considerable frequency dependence of noise exists for {\it long}
conductors ($L \gg \lambda_e, \lambda$).

Despite a considerable recent experimental effort focused on the
shot noise in ballistic\cite{Reznikov 95,Kumar 96} and diffusive\cite{Liefrink
94,Steinbach 96,Schoelkopf 97} structures,  additional experiments are
necessary to determine the exact noise suppression factor and its
dependence on length and frequency.  For wide and short conductors
(like in our first model) such measurements can prove to be extremely
difficult, since in order to implement low internal screening $
(L<\lambda )$ the electron density has to be very low and the sample
very short. However, it seems that the frequency dependence predicted
here could be observed in systems close to our ground-plane model
[Fig. 1(b)], for example in inversion layers of field-effect
transistors and other 2DEG structures.  With typical values of $D'
\simeq D \sim 10^3\,{\rm cm}^2/{\rm s}$ and $L \sim 10 \mu{\rm
m}$\cite{Liefrink 94}, the expected crossover frequency is of the
order of 5 GHz, \ie, within the range currently available for accurate
noise measurements (cf.\ Refs.~\cite {Reznikov 95,Schoelkopf 97}).

To summarize, we have shown that effects of screening have to be taken
into account in order to obtain realistic results for shot noise in
mesoscopic conductors at finite frequencies ($\omega \tau _T\geq
1$). For relatively short conductors ($L\ll \lambda ,\lambda_e$), or
for long, thin conductors close to a ground plane, noise should
approach a level of $(1/2)\times 2eI$ at high frequencies.

We are grateful to M. B\"uttiker for numerous discussions and for
bringing to our attention an error in the initial version of the
manuscript. We are also grateful to K. E. Nagaev and R. J. Schoelkopf
for making their results available prior to publication, to
S. K. Tolpygo for discussions, and to the anonymous referee for a
valuable comment concerning the experimental observation of the
predicted effects. The work was supported in part by DOE's Grant
\#DE-FG02-95ER14575. We thank the Institute for Nuclear Theory at the
University of Washington for its hospitality and the DOE for partial
support during the completion of this work.

\end{multicols}
\end{document}